\documentclass[aip,amsmath,amssymb,reprint]{revtex4-1}
\usepackage[latin1]{inputenc} 
\usepackage[T1]{fontenc}

\usepackage{graphicx}
\usepackage{framed}
\usepackage{esvect}
\usepackage{amsmath,amssymb,color,bm}
\usepackage{mathrsfs}
\usepackage[dvipsnames]{xcolor}

\newcommand\tb[1]{\textbf{#1}}

\newcommand\tn[1]{\textnormal{#1}}

\newcommand\mc[1]{\mathcal{#1}}

\newcommand\rev[1]{\textcolor{black}{#1}}

\newcommand\beq{\begin{equation}}
\newcommand\eeq{\end{equation}}
\newcommand\beqa{\begin{eqnarray}}
\newcommand\eeqa{\end{eqnarray}}

\newcommand\e[1]{\cdot 10^{#1}}
\newcommand\dd{\textnormal{d}}

\def\Z{\mathbb{Z}}

\def\ch{\tn{cosh}}

\def\th{\tn{tanh}}

\def\w{\omega}

\def\v{\tb{v}}

\def\Re{R_\tn{e}}
\def\Ge{G_\tn{e}}
\def\Ri{R_\tn{i}}
\def\Gi{G_\tn{i}}
\def\Zi{Z_\tn{i}}
\def\Rie{R_\tn{ie}}
\def\muie{\mu_\tn{ie}}
\def\Rhe{R_\tn{he}}
\def\muhe{\mu_\tn{he}}
\def\Reo{R_\tn{EO}}
\def\mueo{\mu_\tn{EO}}
\def\Zeo{Z_\tn{EO}}
\def\Rh{R_\tn{h}}
\def\Zh{Z_\tn{h}}
\def\Ue{U_\tn{e}}
\def\Ui{U_\tn{i}}

\def\Ie{I_\tn{e}}
\def\Ii{I_\tn{i}}

\def\C{\mc{C}_{\rm EDL}}
\def\Z{Z_{\rm EDL}}

\def\L{\mc{L}_\tn{h}}

\def\.{\cdot}

\def\1{^{-1}}
\def\2{^{-2}}
\def\3{^{-3}}

\makeatletter
\def\@email#1#2{%
 \endgroup
 \patchcmd{\titleblock@produce}
  {\frontmatter@RRAPformat}
  {\frontmatter@RRAPformat{\produce@RRAP{*#1\href{mailto:#2}{#2}}}\frontmatter@RRAPformat}
  {}{}
}%
\makeatother
\begin{document}

\title{Electron-electrolyte coupling in AC transport through nanofluidic channels}

\author{Baptiste Coquinot}
\affiliation{Laboratoire de Physique de l'\'Ecole Normale Sup\'erieure, ENS, Universit\'e PSL, CNRS, Sorbonne Universit\'e, Universit\'e Paris Cit\'e, 24 rue Lhomond, 75005 Paris, France}
\affiliation{Institute of Science and Technology Austria (ISTA), Am Campus 1, 3400 Klosterneuburg, Austria}

\author{Mathieu Liz\'ee}
\affiliation{Laboratoire de Physique de l'\'Ecole Normale Sup\'erieure, ENS, Universit\'e PSL, CNRS, Sorbonne Universit\'e, Universit\'e Paris Cit\'e, 24 rue Lhomond, 75005 Paris, France}
\affiliation{Present address: Fritz Haber Institute of the Max Planck Society, Faradayweg 4-6, 14195 Berlin, Germany}

\author{Lyd\'eric Bocquet}
\affiliation{Laboratoire de Physique de l'\'Ecole Normale Sup\'erieure, ENS, Universit\'e PSL, CNRS, Sorbonne Universit\'e, Universit\'e Paris Cit\'e, 24 rue Lhomond, 75005 Paris, France}

\author{Nikita Kavokine}
\affiliation{The Quantum Plumbing Lab, \'Ecole Polytechnique F\'ed\'erale de Lausanne (EPFL), Station 6, 1015 Lausanne, Switzerland}

\date{\today}

\begin{abstract}
The transport properties of nanofluidic channels are usually studied under constant (DC) voltage or pressure driving. 
However, the frequency response under sinusoidal (AC) drivings offers rich insights into the time-dependent transport mechanisms.
Inspired by recent electrochemical approaches, we investigate the couplings between ionic and electronic transport under AC driving.
We show that conduction electrons of the channel walls participate in ionic current via capacitive electrochemical coupling, defining a critical frequency and length scale where electron-dominated conductivity emerges.
We further analyze how electron-ion coupling modifies electro-osmotic flows, and demonstrate that fluctuation-induced momentum transfer between the electrolyte and wall electrons produces distinct AC transport signatures depending on the charge carrier polarity. 
Altogether, we establish a  frequency-dependent transport matrix that couples ionic, electronic and hydrodynamic flows. These findings establish  AC nanofluidic transport as a powerful probe of interfacial phenomena under confinement, and suggest new directions for engineering nanofluidic functionalities through electron-electrolyte coupling.
\end{abstract}

\maketitle 

\section{Introduction}

Progress in nanofabrication has recently enabled the study of electrolyte transport through individual fluidic channels with dimensions on the scale of a few molecular sizes~\cite{Faucher2019,Emmerich2024a}. 
Beyond linear electrokinetic (EK) transport, such as electro-osmosis, streaming or osmotic currents~\cite{Siria2013, Feng2016, Emmerich2022}, these nanofluidic devices have unveiled a number of advanced, non-linear EK transport phenomena, such as diode effects, ionic Coulomb blockade, as well as memristive behavior~\cite{Robin2023, Xiong2023a, Bisquert2023, Kamsma2024, Emmerich2024, Barnaveli2024,Vlassiouk2007, Li2023}.
Understanding these phenomena is already central to many applications such as separation processes and energy conversion~\cite{Tristan2020, Logan2012,  Siria2017, Simon2020, Abdelghani-Idrissi2025}. 

Recently, a number of experiments pointed out the role of the channel wall electronic properties in nanoscale ion and fluid transport~\cite{Secchi2016b, Esfandiar2017, Mouterde2019,  Li2022, Goutham2023, Robin2023c, Kondrat2014, Li2024}. 
For example, the screening by conduction electrons has been predicted to accelerate filling dynamics in narrow pores~\cite{Kondrat2014}, while it had no experimentally detectable influence on ion entrance effects~\cite{Li2024}. Concurrently, a systematic difference in ionic conductance has been observed between metallic and semiconducting carbon nanotubes~\cite{cuiCouplingIon2025a}. 
The screening of ionic interactions by conduction electrons in confinement has also been predicted by several theoretical frameworks~\cite{Schlaich2022, Kavokine2022b, Coquinot2023B}. Altogether, nanochannels with electrically conducting (often carbon-based) walls exhibit EK and hydrodynamic responses that are markedly different from those observed in channels with insulating (e.g., boron nitride). However, there is still no clear intuition of how wall electronic properties affect ion transport in nanochannels and membranes~\cite{Xie2024}. 

Experiments have further highlighted the coupling between ionic and electronic pathways for charge transport \rev{under AC voltage} in porous electrodes materials or between graphene layers \cite{Maurel2022,Lizee2025}. The conducting channel walls behave as capacitive electrodes, with the (local) dissolved ion population forming an electric double-layer (EDL), facing its electronic counterpart in the wall. 
This interface is characterized by the corresponding EDL capacitance $\C$ ~\cite{Pireddu2023}, whose value 
ranges between 1 and 50 $\mu$F/cm$^2$~\cite{Simon2008, Miller2010, Uesugi2013}. 
EDL capacitance is the foundation of supercapacitor technology~\cite{Wang2017, Simon2020}, which makes extensive use of nanoconfinement to maximize energy density~\cite{Chmiola2006, Kondrat2016, Salanne2016, Simon2022, Pireddu2024}. When an AC voltage is applied to the nanochannel (Fig. 1a) the EDL capacitor couples the transport of ions in solution to electrons in the channel wall, leading to a "mixed" current carried by both ions and electrons~\cite{Nguyen1999}. While this phenomenon has been discussed in the context of porous materials, 
a thorough framework for the entangled ion and electron transport, including all possible couplings -- see Fig. 1c -- is still missing.

\begin{figure*}
\centering	
\includegraphics[width=\textwidth]{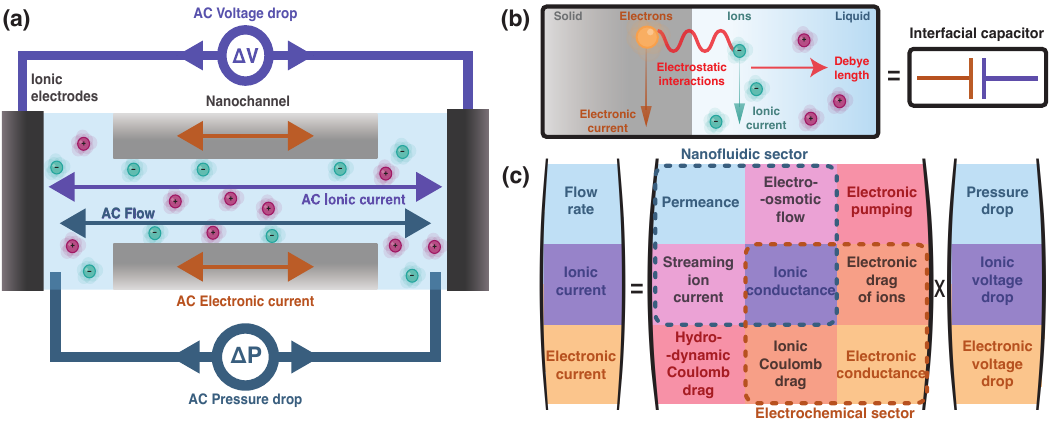}
\caption{\tb{Model}
\tb{(a)} Schematic of the system: an AC ionic current and an AC hydrodynamic flow are driven through a nanochannel with electrically conducting walls by an AC ionic voltage drop and an AC pressure drop.
\tb{(b)} Schematic of the solid-liquid interface: Coulomb interactions couple ions in the electrolyte to electrons in the channel wall.
\tb{(c)} Sketch of the transport matrix of the system. The nanofluidic sector corresponds to the properties investigated in nanofluidic experiments in which we want to integrate the electronic degrees of freedom of the conducting wall. The electrochemical sector corresponds to the properties traditionally investigated in electrochemistry. 
 \label{fig1}}
\end{figure*}

In particular, recent works have shown that
in addition to capacitive effects, there may be direct momentum transfer between an electrolyte and a solid's electrons, due to fluctuation-induced drag effects
~\cite{Schlaich2022, Kavokine2022, Bui2023, Coquinot2023, Coquinot2025}: namely, ionic and hydrodynamic Coulomb drag.
Ionic Coulomb drag originates from the image charges induced by ions in the solid wall. If the wall preferentially adsorbs ions of a given sign, then an ionic current in the electrolyte will result in a net current of image charges~\cite{Rabinowitz2020, Chen2023}.
On the other hand, hydrodynamic Coulomb drag, or hydro-electronic drag, is the solid-liquid analogue of the condensed-matter Coulomb drag effect~\cite{Narozhny2016}. It originates from the quantum or van der Waals liquid-solid friction~\cite{Kavokine2022}, and amounts to momentum transfer between an electrically-neutral liquid and a solid's electrons via interfacial charge fluctuations~\cite{Coquinot2023,Coquinot2024}. 
Recent experiments have evidenced ionic Coulomb drag phenomena~\cite{Rabinowitz2020, Chen2023}, but theoretical predictions suggest that the hydro-electronic drag could be a stronger effect, particularly with carbon-based materials~\cite{Coquinot2024}. 

In this article, we determine the AC transport coefficients of a nanochannel with conducting walls, as sketched in Fig. 1a-b. These can be regrouped into a transport matrix (Fig. 1c), that connects the thermodynamic forces in the system -- hydrostatic pressure drop $\Delta P$, ionic voltage drop $\Delta \Ui$ and electronic voltage drop $\Delta \Ue$ -- to the corresponding fluxes -- electrolyte flow rate $\mc{Q}$, ionic current $\Ii$ and electronic current $\Ie$: 
 \beq 
  \begin{pmatrix}
 \mc{Q} \\ \Ii 	\\  \Ie
  \end{pmatrix}
  \ =
 \begin{pmatrix}
 \L &  \mueo & \muhe  \\ 
\mueo & \Gi &  \muie\\ 
 \muhe & \muie & \Ge
 \end{pmatrix}
\begin{pmatrix}
\Delta P \\ \Delta \Ui  \\  \Delta \Ue
 \end{pmatrix} \label{eq-transport0}
  \eeq
The diagonal terms of the matrix are the hydrodynamic permeance $\L$, the ionic conductance $\Gi$ and the electronic conductance $\Ge$. The cross-terms are the ionic and hydrodynamic Coulomb drag mobilities ($\muie$ and $\muhe$ respectively) and the electro-osmotic mobility ($\mueo$). 
We focus our investigation on how the above-mentioned ion-electron coupling mechanisms modify the transport matrix, in particular under AC driving. 
This requires first coming back to the molecular origin of the various couplings, as is done in
Sec.~\ref{section2}, where we consider the force balance at the ionic-electronic interface. We highlight two regimes of interest: the limit of bulk transport where the ionic and hydrodynamic transport are independent, and the limit of interfacial transport where they are fully intertwined.
In Sec.~\ref{section3}, we focus on the effect of a capacitive electrochemical interface, as sketched in Fig.~1b, calculating the ionic impedance of the channel in the limit of bulk transport.
In particular, we determine conditions for the enhancement of ionic conductivity due to electrons acting as intermediate charge carriers for the ionic current. 
In Sec.~\ref{section4}, we explore the regime of interfacial transport, and investigate how the ionic conductivity and hydrodynamic permeance are modified by the conducting walls. In particular, we study the subtle coupling between electro-osmosis and Coulomb drag effects, which depends on the sign of the solid-state charge carriers, and discuss how the capacitive coupling overall enhances the nanofluidic transport. 

\section{Interfacial transport matrix}\label{section2}

At a molecular level, the solid-liquid interface is the locus of the direct interactions between ions and electrons, as sketched in Fig.1b. The transport of charged species -- be it ions, electrons or holes -- will generate local forces and currents, determined by
local force balances
between the solvent, the ions and the electrons at the mesoscale. The microscopic force balance at the the solid-liquid interface is what allows for the calculation of the transport matrix, and in particular of the cross-terms coming from solvent-ion-electron interactions. 
This description generalizes the modified interfacial stress balance introduced in the presence of a mobile surface charge \cite{Mouterde2018,maduar2015electrohydrodynamics,asmolov2025enhanced}.

As a prototype nanochannel, we consider a nanoscale slit of length $L$, width $W$ and confinement (thickness) $h$, connecting two reservoirs filled with a salt solution (Fig.~1a). We assume that the wall bears a fixed surface charge density $\Sigma$, resulting from the adsorption of ions or the dissociation of surface groups \cite{Mouterde2018}. 
We now add into this well-known description the mobile charge carriers (electrons or holes, with density $n_e$) contained in the wall. 
We assume that their
dynamics are described by a Drude model with intrinsic friction coefficient $\lambda_{\rm e}$~\cite{Coquinot2024}, accounting for the wall's electrical resistance.
A voltage drop and pressure drop (possibly time-dependent) are imposed between the two reservoirs. 

We denote $\v_{\rm h}^{\rm BC}$, $\v_{\rm i}^{\rm BC}$ and $\v_{\rm e}^{\rm BC}$ the resulting interfacial velocities of the liquid, the ions and the electrons, that we assume uniform over the boundary layers where cross-effects originate; 'BC' stands for 'boundary condition', pointing to the interfacial origin of these quantities. 
Here, we will consider only linear transport with constant and uniform coefficients. In particular, we assume that the counterion density at the interface is not affected by variations of the surface potential due to the wall charge carriers, thus neglecting the nonlinear induced charge electro-osmosis effects~\cite{Squires2004}, and potential recirculations of flow.

Let us now write the three momentum balance equations and the resulting fluxes: one for the hydrodynamic flow, one for the (interfacial) ions within the EDL, and one for the electronic charge carriers.
Integrating the momentum balance of the liquid over the whole channel height yields:
\beq \frac{h}{2} \boldsymbol{\nabla} P=\lambda_{\rm h} \v_{\rm h}^{\rm BC} +  \frac{|\Sigma| }{q}\xi_{\rm i}(\v_{\rm h}^{\rm BC}-\v_{\rm i}^{\rm BC}) +  \lambda_{\rm he}(\v_{\rm h}^{\rm BC}-\v_{\rm e}^{\rm BC})\label{eq-transport-liquid} \eeq
where  $\lambda_{\rm h}$ is the usual hydrodynamic friction coefficient on the wall, $\xi_{\rm i}=k_{\rm B}T/D_{\rm i}$ is the Stokes drag coefficient of the ions -- responsible for electro-osmotic coupling -- and $ \lambda_{\rm he}$ is the van der Waals friction coefficient \cite{Kavokine2022} coupling hydrodynamics to the electrons -- giving rise to hydrodynamic Coulomb drag \cite{Coquinot2023, Coquinot2024}. 
Inertial effects have been neglected, which is valid for small pore sizes $h$, and accordingly frequencies below $\nu/h^2$, with $\nu=\eta/\rho$ the kinematic viscosity ($\nu/h^2\sim 10^{12}$ s$^{-1}$ for nanometric confinement).
Eq.~\eqref{eq-transport-liquid} is a generalization of the usual Navier partial slip boundary condition. 
The total flow rate decomposes into a contribution from interfacial slippage and a bulk Hagen-Poiseuille term~\cite{Kavokine2021}:
\beq 
\mc{Q} = hW\v_{\rm h}^{\rm BC} + {h^3W \over 12 \eta }\boldsymbol{\nabla} P
\eeq
where the Poiseuille term accounts for the usual bulk hydrodynamic resistance, acting in parallel with the interfacial contribution.

The momentum balance for the interfacial counter-ions of charge density $-\Sigma$ reads:
\beq -\Sigma \boldsymbol{\nabla} \Ui = \frac{|\Sigma| }{q} \xi_{\rm i}(\v_{\rm i}^{\rm BC} -\v_{\rm h}^{\rm BC} ) +\frac{|\Sigma| }{q} \xi_{\rm e} (\v_{\rm i}^{\rm BC}  - \v_{\rm e}^{\rm BC} ) \label{eq-transport-ions} \eeq 
where $ \xi_{\rm e} $ is the ion-electron drag coefficient~\cite{Persson1995, Liebsch1997}. 
Although typically much smaller than the ion-solvent Stokes friction $\xi_{\rm i}$, we retain this term to explore its physical impact.
The two terms respectively account for electro-osmotic coupling and ionic Coulomb drag.
The resulting ionic current combines an interfacial streaming term and a bulk Ohmic term:
\rev{\beq 
\Ii= -2\Sigma W\v_{\rm i}^{\rm BC} + (2q^2c_s hW D_{\rm i}/k_{\rm B}T)  \boldsymbol{\nabla} \Ui,
\eeq}
where the bulk Ohmic contribution can be estimated from the ionic diffusivity $D_{\rm i}$ and the salt concentration $c_s$ ($q$ is the ionic charge).

Finally, the momentum balance for the wall charge carriers reads:
\beq n_{\rm e}  \boldsymbol{\nabla} \Ue =\lambda_{\rm e}  \v_{\rm e}^{\rm BC}  + \lambda_{\rm he}( \v_{\rm e}^{\rm BC} - \v_{\rm h}^{\rm BC} ) +\frac{|\Sigma| }{q}\xi_{\rm e} ( \v_{\rm e}^{\rm BC} - \v_{\rm i}^{\rm BC} )\label{eq-transport-elec} \eeq
The first term represents the bare electronic resistance used so far. The second and third terms account for Coulomb drag arising from electron-liquid and electron-ion interactions, respectively.
The associated electric current is given by:
\beq
\Ie = 2n_{\rm e} W \v_{\rm e}^{\rm BC},
\eeq
recovering a Drude-like relation for the wall resistance.

These equations can then be rearranged to obtain the matrix of resistances of the system:
 \beq    \begin{pmatrix}
 \boldsymbol{\nabla} P \\    \boldsymbol{\nabla} \Ui  \\   \boldsymbol{\nabla} \Ue
 \end{pmatrix}
 = {1\over L}
 \begin{pmatrix}
 \Rh  &  \Reo & \Rhe  \\ 
\Reo & \Ri &  \Rie \\ 
 \Rhe & \Rie & \Re
 \end{pmatrix}
  \begin{pmatrix}
\mc{Q} \\ \Ii 	\\  \Ie
  \end{pmatrix}\label{eq-transport}
  \eeq
The transport matrix, in Eq.~\eqref{eq-transport0}, is deduced as the inverse of the resistance matrix.
The transport matrix completely describes the various hydrodynamic-ionic-electronic couplings and its general form is quite cumbersome. However,
it takes simple forms
in two limiting cases: bulk and interfacial transport.
This distinction is quantified by a Dukhin number,  $\tn{Du}={\vert \Sigma\vert \over qc_sh}$.

\paragraph{Bulk transport.} Bulk transport is dominating in the limit of a small Dukhin number $\tn{Du}=|\Sigma|/qc_sh\ll1$ corresponding to a small surface charge. 
As a consequence, the ionic current is dominated by its bulk Ohmic contribution and the ionic resistance simply writes:
\rev{\beq \Ri = \frac{k_BT}{2q^2D_{\rm i}c_s}\frac{L}{hW} \quad \tn{and} \quad \Re = \frac{ \lambda_{\rm e}^{\rm eff}L }{2Wn_{\rm e}^2} \label{resistances-bulk}\eeq}
while the electroosmotic and ionic Coulomb drag resistances are negligible. 
In particular, the ionic and hydrodynamic transport are decoupled. 
While the effects of the coupling between the hydrodynamic and the electronic transport due to the hydrodynamic Coulomb drag has been investigated in \cite{Coquinot2024}, we will investigate the effects of the capacitive coupling between the ionic and electronic transport in Sec.~\ref{section3}.

\paragraph{Interfacial transport.} In the limit of a large Dukhin number $\tn{Du}=|\Sigma|/qc_sh\gg1$, interfacial transport is dominating. This corresponds to the limit of a large surface charge; and/or to a small solid-liquid friction (leading to a large slip length $b\gg h$), associated with a uniform velocity profile across the channel.  
As a consequence, the ionic current is dominated by its interfacial streaming contribution and the flow rate by the interfacial slippage.
 Thus, we obtain the hydrodynamic, ionic and electronic resistances
\beq \Rh = \frac{2\eta L}{h^2Wb_{\rm tot}}, \quad  \Ri =  \frac{(\xi_{\rm i}+\xi_{\rm e})L  }{2 qW|\Sigma|}, \quad \Re = \frac{ \lambda_{\rm e}^{\rm eff}L }{2Wn_{\rm e}^2}
\label{resistances}
 \eeq
where $b_{\rm tot}=\eta/(\lambda_{\rm h}+|\Sigma| \xi_{\rm i}/q+\lambda_{\rm he})$ is the total slip length and $ \lambda_{\rm e}^{\rm eff} =  \lambda_{\rm e} + \lambda_{\rm he}+ |\Sigma|\xi_{\rm e}/q$ is the total relaxation rate of the wall charge carriers.
We also obtain the cross-term resistances of electroosmosis, ionic Coulomb drag and hydrodynamic Coulomb drag respectively:
\beq \Reo =  \frac{ \xi_{\rm i}L}{ qhW} \frac{ \Sigma}{|\Sigma| }, \quad \Rie = \frac{\xi_{\rm e}L}{ 2qn_{\rm e} W}  \frac{ \Sigma}{|\Sigma| }, \quad \Rhe = -\frac{\lambda_{\rm he} L }{ n_{\rm e} hW}\eeq
Importantly, we find that the signs of the cross-terms are not universal, as they depend on the relative signs of the various charge carriers.
The cross-term resistances have the opposite sign of the corresponding mobilities.
As the flow drags the counter-ions, the electro-osmotic mobility has the sign of the counterions, and thus $\Reo$ has the sign of the surface charge.
Similarly, the flow drags the wall charge carriers, so that the hydrodynamic Coulomb drag mobility has the sign of $n_{\rm e}$ and $\Rhe$ has the opposite sign. 
Finally, the ionic Coulomb drag pushes the ions and wall charge carriers in the same direction, so that the mobility is positive when they have the same sign, and vice versa for the resistance $\Rie$. For clarity, we assume in the following that the wall charge carriers are mainly negative.
The coupling of this intertwined transport structure with the capacitive interface will be investigated in Sec.~\ref{section4}.

\section{Ionic impedance of a channel with conducting walls}
\label{section3}

\begin{figure*}
\centering	
\includegraphics[width=\textwidth]{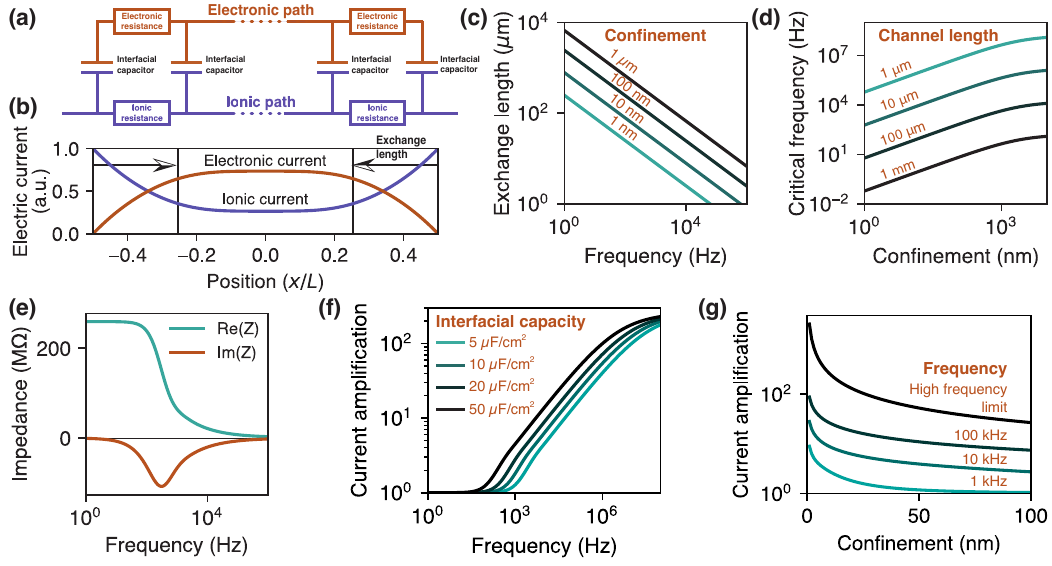}
\caption{\tb{Capacitive effects of the conducting wall on ionic transport.}
 \tb{(a)} Equivalent electronic circuit of the system: the ionic and electronic paths are connected by the continuous interfacial capacitor.
 \tb{(b)} Typical result for the current in the ionic ($J(x)$) and electronic ($1-J(x)$) paths as a function of the normalized position $x/L$ along the channel. The current is exchanged between the two paths over a typical length scale $\ell$, which is indicated by the vertical lines. 
  \tb{(c)} Exchange length $\ell$ as a function of the frequency $\omega/2\pi$ for different confinements $h$. Here, $\C=$ 20 $\mu$F/cm$^2$, $c_s=1$~M, $D_{\rm i}=5\e{-9}$ m$^2$/s and the electronic resistivity is $\Re W/L=$ 10 k$\Omega$.
  \tb{(d)} Critical frequency for opening the electronic path, as a function of the nanochannel confinement $h$ and for different nanochannel lengths $L$. 
\tb{(e)} Real and imaginary parts of the equivalent impedance $Z$ of the system as a function of frequency, assuming a nanochannel length $L=100$ $\mu$m, confinement $h=$ 10 nm, interfacial capacity $\C=$ 20 $\mu$F/cm$^2$, salt concentration $c_s=1$ M, ionic diffusivity  $D_{\rm i}=5\e{-9}$ m$^2$/s and electronic resistivity $\Re W/L=$ 10 k$\Omega$. 
  \tb{(f)} Current amplification, defined as the ratio of the electric current in the presence and in the absence of the electronic path, as a function of the frequency, for varying interfacial capacities $\C$. 
   \tb{(g)} Current amplification as a function of the confinement $h$, for varying frequencies as well as the high frequency limit. 
 \label{fig2}}
\end{figure*}

First, we investigate the consequences of the ion-electron capacitive coupling in the absence of surface charge.
Thus, we focus on the electrochemical part of the transport matrix in Fig.~1c, where the ion-wall coupling is characterized by the double-layer capacitance $\C$ -- corresponding to \rev{a total impedance $\Z = 1/(i \omega\C WL)$} -- and neglect all transport cross-effects.
Our goal is to determine the ionic current and flow rate under AC forcing, accounting for the presence of conduction electrons in the walls.
Under the assumption that electrostatic equilibration across the channel is faster than the variation of the applied potential, the channel's impedance can be described by a modified transmission line model (TLM) -- see Fig.~2a. 
It is made of two resistive paths -- ionic and electronic (the latter representing the channel wall), connected by the interfacial capacitor. 
Thus, the difference between the ionic and electronic potential is given by the accumulation of charges in this capacitor:
  \beq \Ui(x)-\Ue(x)=\Z L\partial_x \Ii(x) \label{eq-EDL}\eeq
  In the absence of Coulomb drag or electroosmotic effects, the spatial evolution of these potentials is given by the Ohm's laws
\beq L\partial_x \Ui(x) = \Ri \Ii(x)  \quad \tn{and}  \quad L\partial_x \Ue(x) = \Re \Ie(x) \label{eq-Ohm}\eeq
Only the ionic path is connected to the external circuit, but the electronic path can contribute to the AC conductivity thanks to the interfacial capacity. 
We note that the complete circuit comprises a resistance and a capacitance due to the electrodes and the solution reservoirs in series with the nanochannel. 
In the following, we focus only on the nanochannel impedance $\Zi(\omega)= \Delta \Ui / \Ii$ where $ \Delta \Ui $ is the total ionic potential drop along the nanochannel, keeping in mind that the electrodes impose a cutoff for the frequencies $\omega$ that can be probed, typically ranging from 1 kHz to 1 MHz.

If the two paths were perfectly connected, the fraction of electric current going through the ionic path would be $I_{\rm i}^0/I=\Re/(\Ri+\Re)$ where $I$ is the total current going through the transmission line. 
But with a non-trivial interfacial impedance, this fraction becomes dependent on the position $x$ along the channel. 
Gathering Eqs.~\eqref{eq-EDL}-\eqref{eq-Ohm} and using that the total current $I=\Ii+\Ie$ is conserved, we find that the excess ionic current $J(x)=I_{\rm i}(x)-I_{\rm i}^0$ solves a modified Telegrapher's equation in the continuous limit of the TLM~\cite{Nguyen1999}:
\beq \ell(\omega)^2\partial_x^2 J(x,\omega)= 2iJ(x,\omega)\label{telegrapher1}\eeq	
where we have introduced an exchange length: 
\beq \ell(\omega)=\sqrt{\frac{2L}{\omega\C W(\Ri+\Re)}}.\label{ell}\eeq
Here, this length has been defined to be a real value, and the factor $i$ of the capacitive impedance $\Z$ has been explicitly written in Eq.~\eqref{telegrapher1}.
Physically, $\ell$ corresponds to the lengthscale over which current is exchanged between the two paths, as shown in Fig.~2b where a typical solution of Eq.~\eqref{telegrapher1} is plotted. At distances $x \gtrsim \ell(\omega)$ along the channel, $J(x) = 0$ so that $I_{\rm i} = I_0$. Since the electronic resistance is typically much lower than the ionic resistance, this means that electrons become the main charge carriers far enough from the channel mouth. This transition is of course only possible if the channel length $L > \ell(\omega)$ -- a condition easiest to satisfy at higher frequencies. 
At a given frequency, $\ell(\omega)$ is shorter for stronger confinement: the larger the ionic resistance, the easier the transition to electronic conduction (Fig.~2c).

We confirm this qualitative intuition by solving for the impedance $\Zi(\omega)$ of the channel using the boundary conditions $I_{\rm i}(\pm L/2)=I$.
Indeed, the Telegrapher's equation provides the ionic current profile given the total total current $I$:
    \beq \Ii(x)=\Ii^0+(I-\Ii^0)\frac{\ch((1+i)x/\ell)}{\ch((1+i)L/2\ell)}\eeq
    Then, integrating the Ohm's law leads to the total ionic potential drop:
    \beq \Delta \Ui = \Ri\int_{-L/2}^{L/2}\frac{\dd x}{L} \Ii(x) \eeq
    and ultimately to the impedance:
\beq \Zi(\omega)= \frac{\Re\Ri}{\Re+\Ri}\left[1+\frac{ \Ri}{\Re}\Phi\left(\frac{\omega}{\omega_c}\right) \right] \label{eq-Zi}\eeq
which is consistent with the result of~\cite{Nguyen1999} for porous media.
Here, we have introduced a critical frequency $\omega_c$, defined by $\ell (\omega_c) = L$, which reads explicitly (Fig.~2d):
\beq \omega_c= \frac{2}{\C W L(\Ri+\Re)}\eeq
and a complex low-pass filter function $\Phi(x)=\th\left(\sqrt{2ix}\right)/\sqrt{2ix}$ which goes from 1 for $x\ll 1$ to 0 for $x\gg1$. 
\begin{itemize}
\item At $\omega = 0$, $Z = \Ri$: the electronic path cannot contribute to the DC conductivity because the electrons cannot exit the channel wall \rev{and the EDL capacitance blocks DC currents.}
\item At $\omega \to \infty$, $Z \approx \Re$: the electronic path effectively short-circuits the ionic path \rev{as the EDL impedance vanishes.} $\omega = \omega_c$ marks the transition between these two regimes. At this frequency, the imaginary part of the impedance is largest and the channel exhibits capacitive behavior (see Fig.~2e-f). 
$\omega_c^{-1}$ also replaces the ionic diffusive time as the time of electric transport through the channel.
\end{itemize}
These two limits were indeed highlighted in the experiments of Refs. \cite{Maurel2022,Lizee2025}.
In practice, the electronic path is relevant if $\omega_c$ is within the frequency range accessible to the ionic electrodes, typically between DC and $10^3 - 10^6~\rm Hz$. As shown in Fig.~\ref{fig2}d, this is the case for channels with high aspect ratio, 
\rev{but not for microscopic drops like in \cite{Lizee2025}.}
For realistic values of the resistivities and interfacial capacitance (see Fig. 1 legend), we find for instance $\omega_c = 10~\rm Hz$ for $L = 100~\rm \mu m$ at 1 nm confinement, or $L = 1~\rm mm$ at $1~\rm \mu m$ confinement. Electron-mediated ionic conduction can thus occur across a vast range of scales, from nanochannels to microchannels. 
We note, however, that there is a significant effect on the overall conductivity only if the electronic resistance is much lower than the ionic resistance, which is not necessarily the case for larger channels (Fig.~2g).

\begin{figure}
\centering	
\includegraphics[]{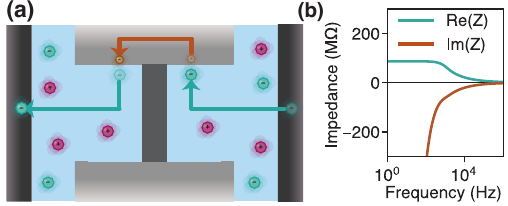}
\caption{\tb{AC conduction through a clogged channel.}
\tb{(a)} Schematic of a clogged channel, where direct ionic conduction between the two ionic electrodes is blocked. The main path taken by the AC current is represented.
 \tb{(b)} Impedance of the system with the same parameters as in Fig. \ref{fig2}.  
 \label{fig3}}
\end{figure}

AC conductivity measurements thus appear as sensitive probes of confined electrode-electrolyte interfaces. Such measurements can also provide valuable insight into multi-channel systems. Indeed, at frequencies higher than $\omega_c$, ions are no longer crossing the channel, meaning that even a clogged channel should exhibit non-zero conductivity. 
In this scenario, the ions connect the electrode to the interfacial capacitance of the nanochannel, while the electrons connect the two sides of the nanochannel. 
A clogged channel as in Fig. 3a is modeled electrically by two modified TLMs (Fig.~2a) in series, with alternating ionic and electronic outputs. 
In particular, Eq.~\eqref{telegrapher1} remains valid with adapted boundary conditions.
The voltage drop applied on each side of the system now writes:
\beq \frac{\Delta \Ui}{2}=\frac{\rho_{\rm i}}{Wh}\int_{-L/2}^0\dd x\, I_{\rm i}(x)-\Z\partial_x I_{\rm i}(0),\eeq
leading to an equivalent impedance (see Fig. 3b):
\beq \Zi =\frac{\Ri\Re}{\Ri+\Re}+\frac{1}{4}\frac{(\Ri-\Re)^2}{\Ri+\Re}
\Phi\left(\frac{\omega}{4\omega_c}\right) +
\frac{\Ri+\Re}{\Phi\left(\frac{\omega}{4\omega_c}\right)}
\label{impedance_2}
\eeq
Here, at low frequencies, the last term diverges: no current can cross the circuit since there is no exchange between the two paths. 
Such a term is obtained in the modelling of supercapacitors in~\cite{Maurel2022}.
However, at larger frequencies, electrons carry the current through the channel, so that the total resistance is again the one of the parallel circuit $\Re\Ri/(\Re+\Re)$.  
Impedance spectroscopy can thus be used to probe the number of clogged channels -- currently an experimental challenge~\cite{Holt2006, Emmerich2024a} -- in nanoporous membranes or multichannel nanofluidic systems with conducting walls. 

\section{Interfacial nanofluidic transport through a conducting nanochannel}
\label{section4}

\begin{figure*}
\centering	
\includegraphics[width=\textwidth]{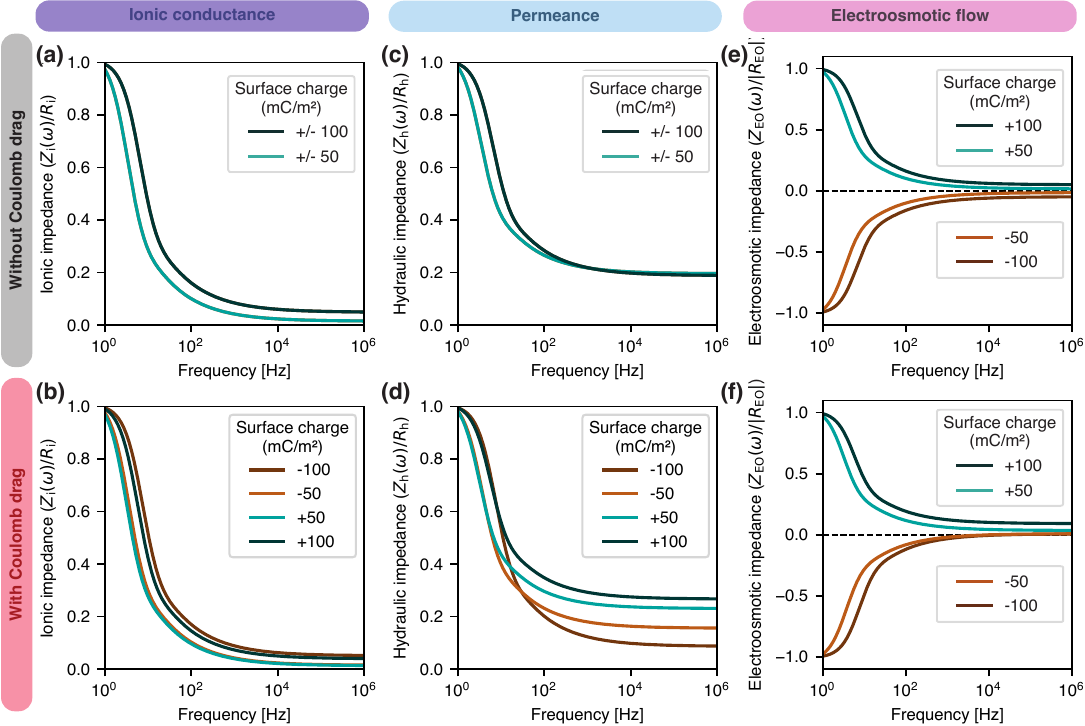}
\caption{\tb{Nanofluidic impedances with conducting walls.} 
We consider a channel where transport is dominated by surface charge (Du $\gg 1$) and slippage ($b\gg h$), of height $h=10$ nm, width $W= 1 \mu$m and length $L=100 \mu$m, with interfacial capacity $\C=$ 20 $\mu$F/cm$^2$, ionic diffusivity  $D_{\rm i}=5\e{-9}$ m$^2$/s, electronic friction $\xi_{\rm e}=10^{-13}$ kg/s, bare hydrodynamic friction coefficient $\lambda_{\rm h} = 10^{4}$ Pa.s/m, van der Waals friction coefficient $\lambda_{\rm he}=10^4$ Pa.s/m and a solid with a charge density $n_{\rm e}=-10^{14}e$/cm$^2$ and a bare electronic resistivity $\Re W/L=$ 10 k$\Omega$ \cite{Coquinot2024}.
 \tb{(a)-(b)} Ionic impedance $\Zi$ normalized by its DC value $\Ri$ as a function of frequency for various surface charges.
  \tb{(c)(d)} Hydrodynamic impedance $\Zh$ normalized by its DC value $\Rh$ as a function of frequency for various surface charges. 
  \tb{(e)-(f)} Electro-osmotic impedance $\Zeo$ normalized by its DC value $|\Reo|$ as a function of frequency for various surface charges. 
We have set Coulomb drag cross-terms to zero ($\Rie=\Rhe=0$) in the first row and restored them in the second row of plots. 
 \label{fig4}}
\end{figure*}

\begin{figure*}
\centering	
\includegraphics[width=\textwidth]{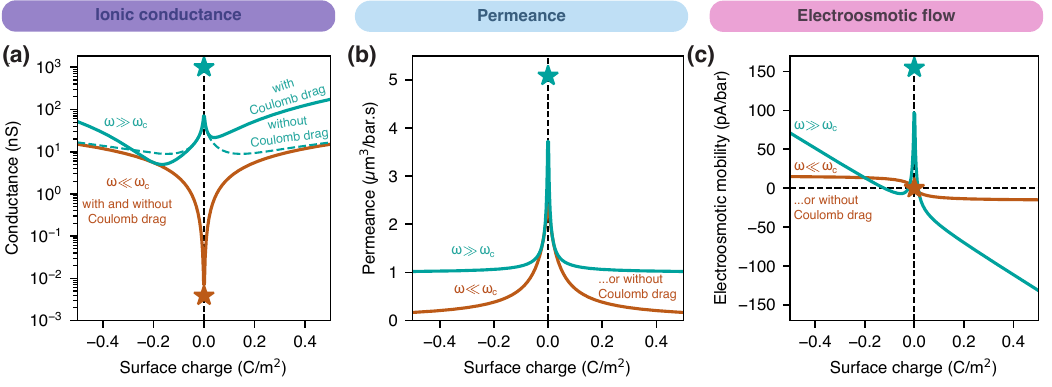}
\caption{\tb{Nanofluidic transport with conducting walls.} Same system parameters than in Fig.~\ref{fig4}.
    \tb{(a)} Low and high frequency ionic conductance $G_{\rm i}$ of the channel as a function of the surface charge. 
    In the absence of Coulomb drag ($\Rie=\Rhe=0$) the conductance is identical at low frequency and corresponds to the dashed green line at high frequency.
 \tb{(b)} Low and high frequency permeance $\L$ of the channel as a function of the surface charge.
\tb{(c)} Low and high frequency electro-osmotic mobility $\mu_{\rm EO}$ of the channel as a function of the surface charge. 
       For \tb{(b)-(c)} the transport coefficient is independent of frequency in the absence of Coulomb drag and corresponds to the low-frequency limit. 
          The stars indicate the low surface charge limit obtained from the bulk transport resistances in Eq. \eqref{resistances-bulk}, with $c_s= 1$ mM, and including the hydrodynamic Coulomb drag in \tb{(c)}. 
 \label{fig5}}
\end{figure*}

We now turn to the case of interfacial transport where the cross-terms of the transport matrix (see Fig.~\ref{fig1}c) can further couple the electronic transport with the ionic and hydrodynamic transport.
We start again from Eq.~\eqref{eq-EDL} for the interfacial capacity:
  \beq \Ui(x)-\Ue(x)=\Z L\partial_x \Ii(x) \label{eq-EDL2}\eeq
  that we differentiate, and replace the Ohm's laws Eqs.~\eqref{eq-Ohm} by the generic matrix of resistances Eq.~\eqref{eq-transport}:
  \beq    \begin{pmatrix}
 \boldsymbol{\nabla} P \\    \boldsymbol{\nabla} \Ui  \\   \boldsymbol{\nabla} \Ue
 \end{pmatrix}
 ={1\over L}
 \begin{pmatrix}
 \Rh  &  \Reo & \Rhe  \\ 
\Reo & \Ri &  \Rie \\ 
 \Rhe & \Rie & \Re
 \end{pmatrix}
  \begin{pmatrix}
\mc{Q} \\ \Ii 	\\  \Ie
  \end{pmatrix}\label{eq-transport2}
  \eeq
 Thus, using conservation of the total curent, we obtain the general equation for the ionic current as:
  \beqa \Z L^2\partial_x^2 \Ii(x) &=& (\Re+\Ri-2\Rie)\Ii(x)-(\Re-\Rie)I \nonumber\\
  && \ldots+(\Reo-\Rhe)\mc{Q}\eeqa whose uniform solution is now:
    \beq \Ii^0=\frac{\Re-\Rie}{\Re+\Ri-2\Rie}I+\frac{\Rhe-\Reo}{\Re+\Ri-2\Rie}\mc{Q}\eeq
    Note that the denominators are positive due to the positivity of the matrix of resistances. 
     Finally, the excess ionic current $J(x)=I_{\rm i}(x)-I_{\rm i}^0$ solves the same Telegrapher's equation as before:
        \beq \ell(\omega)^2\partial_x^2 J(x,\omega)= 2iJ(x,\omega)\label{telegrapher2}\eeq
         with the modified exchange length
         \beq \ell(\omega)=\sqrt{\frac{2L}{\omega\C W(\Ri+\Re-2\Rie)}}\label{ell2}\eeq
         which can be either enhanced or reduced compared to the surface-charge-free case, depending on the sign of $\Rie$. Let us still define the critical frequency $\omega_c$ such that $\ell(\omega_c)=L$.

      Overall, using the boundary conditions $I_{\rm i}(\pm L/2)=I$, we then obtain the ionic current profile:
   \beqa
      \Ii(x)&=&\left[\frac{(\Re-\Rie) + (\Ri-\Rie)\frac{\ch\left(\frac{(1+i)x}{\ell}\right)}{\ch\left(\frac{(1+i)L}{2\ell}\right)}}{\Re+\Ri-2\Rie}\right] I \\
    && \ldots +\frac{\Rhe-\Reo}{\Re+\Ri-2\Rie} \left[1-\frac{\ch\left(\frac{(1+i)x}{\ell}\right)}{\ch\left(\frac{(1+i)L}{2\ell}\right)}\right]\mc{Q}\nonumber
     \eeqa
     Finally, we make use of Eq.~\eqref{eq-transport} to obtain the total ionic potential drop:
      \beq \Delta \Ui = (\Ri-\Rie)\int \frac{\dd x}{L}\Ii(x) +\Rie I+\Reo\mc{Q} \eeq
      and the total pressure drop:
        \beq \Delta P = (\Reo-\Rhe)\int \frac{\dd x}{L}\Ii(x) +\Rhe I+\Rh \mc{Q}  \eeq
    Hence, we deduce an effective transport matrix for the nanochannel that accounts for the presence of the conducting walls:
    \beq \begin{pmatrix}
 	   \Delta P (\w) \\  \Delta\Ui (\w)
 \end{pmatrix}
 =
 \begin{pmatrix}
  \Zh(\w) & \Zeo(\w) \\
 \Zeo(\w) & \Zi(\w)
 \end{pmatrix}
  \begin{pmatrix}
\mc{Q} (\w) \\  \Ii (\w) 
  \end{pmatrix} \label{eq-transport2}
  \eeq
  These effective ionic, hydraulic and electro-osmotic impedances determine the linear response of the system to a dynamical (AC) voltage and pressure drop, which can be probed experimentally. 
  
  The effective ionic impedance reads
  \beq \Zi(\w) = \Ri-\frac{(\Ri-\Rie)^2}{\Re+\Ri-2\Rie} \left[1-\Phi\left(\frac{\omega}{\omega_c}\right)\right] \eeq
  and is plotted in Fig.~\ref{fig4}a-b with and without($\Rie=\Rhe=0$) Coulomb drag effects.
 We observe that the impedance suppression at high frequencies is weaker with increasing surface charge: more surface charge means more counterions to carry the electric current (lower DC ionic resistance), hence less transport through the electronic path.   
 We further observe in Fig.~\ref{fig4}b that including the ionic Coulomb drag effect further affects the impedance in a way that depends on the sign of the surface charge. If the surface charge is positive, then the charge carriers in the solution and in the wall are both negative, and ionic Coulomb drag favors current exchange between the ionic and electronic paths ($\Rie < 0$) -- hence a stronger impedance reduction. If the surface charge is negative, then the charge carriers are of opposite signs, $\Rie > 0$, and the drag effect hinders ion transport, thus increasing the impedance. 

The effective hydraulic impedance reads:
\beq \Zh(\w) =  \Rh-\frac{(\Rhe-\Reo)^2}{\Re+\Ri-2\Rie} \left[1-\Phi\left(\frac{\omega}{\omega_c}\right)\right] \eeq
and is plotted in Fig.~\ref{fig4}c-d. 
Similarly to the ionic impedance, the hydraulic impedance is suppressed at high frequencies as a consequence of the current exchange between the electronic and ionic paths -- which impacts the flow through electro-osmosis.  
 Hydrodynamic Coulomb drag further suppresses the impedance if the ionic and electronic charge carriers are of different signs, that is if the surface charge is negative (Fig.~\ref{fig4}d). On the contrary, Coulomb drag increases the impedance in the case of positive surface charge.  

Finally, the electro-osmotic impedance reads:
 \beq \Zeo(\w) =\Reo -\frac{(\Ri-\Rie)(\Reo-\Rhe)}{\Re+\Ri-2\Rie} \left[1-\Phi\left(\frac{\omega}{\omega_c}\right)\right] \label{eq-Zeo} \eeq
and is plotted in Fig.~\ref{fig4}e-f.
 At low frequency, we recover the electro-osmotic resistance $\Reo$ whose sign depends on the sign of the surface charge. At high frequency, electro-osmosis all but vanishes: indeed, the coupling between liquid flow and ionic current stems from the physical motion of the ions, that drag the liquid. If the ionic current is being carried through the electronic path, then the coupling mechanism disappears. Hydrodynamic Coulomb drag re-introduces a coupling mechanism that allows for non-zero electro-osmosis at high frequencies (Fig.~\ref{fig4}f); the corresponding contribution to the impedance is always positive, since the electrons are negative -- regardless of the sign of the surface charge. 

While the matrix of resistances discussed above illuminates the intricate physics of fluid-ion-electron couplings, the experimental observable is rather its inverse, the transport matrix. Indeed, we usually have control over the forces (we may set the pressure or the voltage drop to zero) but not over the fluxes (it is harder to impose zero current or flow rate). 
We obtain the transport matrix by inverting Eq.~\eqref{eq-transport2}: 
    \beq 
 \begin{pmatrix}
\mc{Q} (\w) \\  \Ii (\w) 
  \end{pmatrix} =
 \begin{pmatrix}
  \L(\w) & \mueo(\w) \\
 \mueo(\w) & \Gi(\w)
 \end{pmatrix}
 \begin{pmatrix}
 	   \Delta P (\w) \\  \Delta\Ui (\w)
 \end{pmatrix}
  \label{eq-transport3}
  \eeq
The transport coefficients contain the same physics as the impedances, but their qualitative behavior is quite intricate
and deserves a separate discussion.

The ionic conductance reads:
\beq  \Gi(\w) = \frac{1/\Zi(\w) }{1- \Zeo(\w)^2/ (\Zi(\w)\Zh(\w)) }\eeq
Fig.~\ref{fig5}a shows its low and high frequency limits.
At low frequency, the conductance is dominated by electrophoresis rather than electro-osmosis with our choice of parameters, so that $\Gi(0)\propto |\Sigma|$, as expected for a channel where the ion concentration is imposed by the surface charge. Note that our model thus reaches its limit of validity when $|\Sigma| \lesssim c_s h$, where $c_s$ is the bulk salt concentration. At high frequency, the conductance is boosted by the electronic path and further boosted by Coulomb drag effects, the latter being dependent on the sign of the surface charge. The conductance boost is stronger when ionic and electronic charge carriers are of the same sign, that is when the surface charge is positive. 
 
 The hydraulic permeance reads:
 \beq  \L(\w) = \frac{1/\Zh(\w) }{1- \Zeo(\w)^2/(\Zi(\w)\Zh(\w)) }\eeq
Fig.~\ref{fig5}b shows its low and high frequency limits. While the hydraulic impedance shows a strong frequency dependence (Fig. 4c-d), the permeance is independent of frequency in the absence of Coulomb drag effects. Indeed, if we impose zero ionic (streaming) potential, then the ions are prohibited from moving whatever the frequency of the pressure forcing, so that the permeance scales with the total slip length $\L\propto b_{\rm tot}$, which includes the friction of the flow on the ions: $b_{\rm tot}=\eta/(\lambda_{\rm h}+|\Sigma| \xi_{\rm i}/q+\lambda_{\rm he})$. Since we work with large surface charges, we have approximately $\L(\omega \rightarrow 0)\propto1/|\Sigma|$; note here as well that $|\Sigma| \lesssim c_s h$ is beyond our model's limit of validity. Now, if we allow for hydrodynamic Coulomb drag, then under AC forcing the electrons follow the liquid flow and thus enhance the permeability; since it does not involve the ions, this effect does not depend on the sign of the surface charge. 

 Finally, the electro-osmotic mobility reads:
  \beq  \mueo(\w) = -\frac{\Zeo(\w) }{\Zi(\w)\Zh(\w)- \Zeo(\w)^2}\eeq
 Fig.~\ref{fig5}c shows its low and high frequency limits, which are very different. 
At low frequency or in the absence of Coulomb drag effects, the electro-osmotic flow saturates for large surface charges, since the liquid-ion friction imposes $\Zh(\w)\propto |\Sigma|$: the ions drive the flow but also increase the hydrodynamic resistance. 
 Nevertheless, thanks to the hydrodynamic Coulomb drag combined with the interfacial capacitance, the conducting walls allow for bypassing this saturation at high frequency as the ionic current is replaced by an electronic current.
 Hence, at high frequency, we predict $\mueo\propto |\Sigma|$, with details of the low $|\Sigma|$ behavior dependent on the relative signs of the charge carriers. Notably, at zero surface charge, we find a non-vanishing AC electro-osmotic mobility (see the star in Fig. 5c). Indeed, even without net ionic charge in the channel, part of the ionic current is being carried through the electronic path, and the electronic current induces a liquid flow through hydrodynamic Coulomb drag. 

Two general trends thus emerge from the intricacies of AC the transport coefficients. First, electrolyte-electron coupling (coming from either capacitive or drag effects) results in enhanced ion and fluid transport at high frequencies. This enhancement is due to electrons dissipating less energy than ions in transporting a given amount of charge. Second, Coulomb drag effects introduce a dependence of the transport coefficients on the relative sign of ionic and electronic charge carriers -- a valuable feature for determining the nature of the surface charge in a nanofluidic system. 

\section{Conclusion}
\label{section-conclusion}

We have introduced a general framework describing the intertwined ionic and electronic pathways for nanofluidic charge transport under a sinusoidal (AC) driving.
We predict that nanofluidic channels with conducting walls exhibit enhanced AC conductivity, as electrons in the walls act as intermediate charge carriers for the ionic current via the interfacial capacitance. Through interfacial cross-coupling effects, this enhancement propagates to an increased hydraulic permeability and electro-osmotic mobility. 
Beyond single-channel systems of fundamental interest, this enhanced transport may find  applications in filtration or energy harvesting processes that exploit AC forcing~\cite{Siria2017, Abdelghani-Idrissi2025, Chapuis2025}, that could therefore benefit from the use of conducting membranes, such as conducting polymers~\cite{Chiang1977, Heeger2001} or MXenes~\cite{Gogotsi2019}.
The effect would be particularly significant in long channels, such as those encountered in tortuous porous media~\cite{Zhang2020, Ghanbari2020, Yang2024}. 
This is therefore a promising avenue for enhancing the efficiency of osmotic energy conversion -- for instance, the reverse electrodialysis process -- by shortening transient regimes~\cite{Chapuis2025}, or via enhanced electro-osmotic transport~\cite{Abdelghani-Idrissi2025}.

Fundamentally, our findings also highlight the rich phenomenology unveiled by using an AC probe and points to the potential of impedance spectroscopy and noise spectroscopy~\cite{Robin2023c} as sensitive probes of confined interfacial phenomena~\cite{Lizee2025}. 
Our theoretical framework enables a detailed interpretation of frequency-dependent impedances in terms of electrostatic processes under nanoconfinement, and allows one to disentangle electro-osmosis from ionic and hydrodynamic Coulomb drag effects. 
While ionic electrodes already allow for current measurements up to the MHz range -- especially in their capacitive limit -- experimental techniques for high-frequency hydrodynamic impedance spectroscopy are still lacking. 
Bridging nanofluidics with acoustics, by studying wave propagation through nanochannels where the hydrodynamic impedance plays the role of a damping coefficient, appears as a promising route to overcome this limitation.
 Then, these probes may inform on ionic conductance, interactions and screening under strong confinement, at the heart of outstanding fundamental questions in the fields of energy storage~\cite{Chmiola2006, Kondrat2016, Salanne2016, Simon2022}, molecular separation~\cite{Han2007,  Esfandiar2017, Fumagalli2018} and electrochemical CO$_2$ reduction~\cite{Shin2022, Vavra2024}.

Future work will extend our modeling to nanochannels with more general electrochemical (Faradaic) interfaces. We anticipate that such systems may host rich and subtle functionalities. For instance, applying an AC voltage to the electronic path instead of the ionic one could induce ion transport against a concentration gradient, through a mechanism reminiscent of redox-powered biological proton pumps~\cite{Kim2007}.

\section*{Acknowledgments}
The authors thank Nicolas Chapuis for fruitful discussions. L.B. acknowledges support from ERC project {\it n-AQUA}, grant agreement $101071937$. B.C. acknowledges support from the CFM Foundation and the NOMIS Foundation. N.K. acknowledges support from the Swiss National Science Foundation (SNSF), grant number CRSK-2\_237930. 

\section*{Data availability}

No data was generated in this study.

\end{document}